\documentclass[aps,prc,twocolumn,superscriptaddress,showpacs]{revtex4}

\usepackage{epsfig}
\usepackage{graphicx}
\usepackage{fancyhdr}
\usepackage{pslatex}   
\usepackage{color}
\usepackage{amsmath, amsthm, amssymb} 
\usepackage{natbib}
\usepackage{multirow}

\newcommand {\UrQMD} {UrQMD}

\begin{document}

\title{Correspondence between HBT radii and the emission zone in
non-central heavy ion collisions.}




\author{E. Mount}
  \affiliation{Department of Physics, Ohio State University, Columbus, Ohio 43210, USA}
\author{G. Graef}
  \affiliation{Frankfurt Institute for Advanced Studies, Frankfurt am Main, Germany}
  \affiliation{Institut f\"{ur} Theoretische Physik, Goethe-Universit\"{a}t, Frankfurt am Main, Germany}
\author{M. Mitrovski}
  \affiliation{Frankfurt Institute for Advanced Studies, Frankfurt am Main, Germany}
\author{M. Bleicher}
  \affiliation{Frankfurt Institute for Advanced Studies, Frankfurt am Main, Germany}
  \affiliation{Institut f\"{ur} Theoretische Physik, Goethe-Universit\"{a}t, Frankfurt am Main, Germany}
\author{M.A. Lisa}
  \affiliation{Department of Physics, Ohio State University, Columbus, Ohio 43210, USA}




\begin{abstract}
In non-central collisions between ultra-relativistic heavy ions, the freeze-out distribution
  is anisotropic, and its major longitudinal axis may be tilted away from the beam direction.
The shape and orientation of this distribution are particularly interesting, as they provide
  a snapshot of the evolving source and reflect the space-time aspect of anisotropic flow.
Experimentally, this information is extracted by measuring pion HBT radii as a function of angle
  with respect to the reaction plane.
Existing formulae relating the oscillations of the radii and the freezeout anisotropy are in
  principle only valid for Gaussian sources with no collective flow.
With a realistic transport model of the collision, which generates flow and non-Gaussian sources,
  we find that these formulae approximately reflect the anisotropy of the freezeout distribution.
\end{abstract}

\maketitle

\section{INTRODUCTION}

Femtoscopic intensity interferometry measurements use two-particle correlation functions to probe the
  space-time substructure of the system generated in heavy ion collisions, at the femtometer scale.
At relativistic collision energies, extensive systematic studies have mapped
  out this substructure as a function of transverse momentum ($p_T$), rapidity ($y$), collision
  energy  ($\sqrt{s_{NN}}$) and the mass of the two correlated particles; see~\cite{Lisa:2005dd} for a recent review.
These measurements probe the space-time geometry of the freezeout distribution-- the distribution of
  last-scattering points of the particles.
In addition to the source size and lifetime at freezeout, the momentum-dependence of the femtoscopic scales
  reveal the coordinate-space aspects of collective motion-- ``flow.''

It has long been recognized that measurements relative to the direction of the impact
  parameter of the collision are more sensitive to important underlying physics of the system,
  than are angle-integrated measurements.
The azimuthal dependence of particle yields and spectra-- often called ``directed'' and ``elliptical flow''--
  are extensively used to extract the QCD equation of state (EoS) and transport coefficients of the
  quark-gluon plasma such as viscosity~\cite{Kolb:2000sd,Voloshin:2008dg}.
On the other hand, azimuthally-integrated $p_T$ spectra can flag the existence of collective behavior, but are
  not as discriminating between different models producing such behavior~\citep[e.g.][]{Lisa:1994yr,Song:2010mg}.
Similarly, the azimuthal dependence of jet quenching is a more discriminating probe of partonic
  energy loss, than azimuthally-integrated measurements~\cite{Cole:2005yv,Majumder:2006we,Bass:2008rv}.

It is possible that azimuthally-differential analysis might yield a similar improvement in sensitivity
  of femtoscopy.
Only a few measurements of femtoscopic pion correlations relative to the impact parameter have been
  reported~\cite{Lisa:2000xj,Adams:2003ra,Adamova:2008hs}, though an extensive energy-dependence of these measurements
  is underway at RHIC~\cite{Aggarwal:2010cw}.

Most femtoscopic analyses of pion correlations extract so-called ``HBT radii'' (c.f. discussion in~\cite{Lisa:2005dd})
  which fully describe the emission distribution only if it is purely Gaussian and features no collective motion.
Strictly speaking, neither of these conditions characterize heavy ion collisions, and there has been considerable activity
  in measuring the non-Gaussian features of the source via ``imaging''
  techniques~\cite{Brown:1997ku,Panitkin:1999yj} that fit the source with a sum of spline functions.
Large non-Gaussian tails are mostly explained by long-lived resonance production~\cite{Brown:2007raa}.

Gaussian HBT radii seek to capture the bulk scales of particular interest; several studies exist, testing the
  correspondence between HBT radii and the source scales of non-Gaussian, flowing distributions from
  cascade calculations~\cite{Hardtke:1999vf,Lin:2002gc} and
  blast-wave and hydrodynamical models~\cite{Frodermann:2006sp}.


Moving beyond HBT radii themselves, which characterize the geometry only of subsets
  (``homogeneity regions''~\cite{Makhlin:1987gm}) of the overall source, one is interested
  in the shape and orientation of the emission region as a whole.
Retiere and Lisa have propsed~\cite{Retiere:2003kf} a formula connecting the azimuthal oscillations of HBT radii
  with the transverse anisotropic shape of the source.
Here, we follow the same line and propose a formula for the tilt of the source relative to the beam
  direction.
Both of these formulae are strictly valid only for Gaussian, non-flowing sources.
In this paper, we test these formulae in the context of a realistic transport model featuring non-Gaussian
  freezeout distributions and strong flow.
We find reasonable consistency between the source eccentricity and tilt as extracted directly from the
  space-time freezeout coordinates, and the same quantities estimated with the formulae.
The discrepancy between the two provides an estimate of the systematic uncertainty one expects,
  when using azimuthally-differential pion correlation measurements to extract the underlying source
  shape and orientation.

In section~\ref{sec:formalism}, we define the formalism and the connection between
  HBT radii and an anisotropic but simplified static Gaussian source.
There, we present formulae connecting measurable quantities
  to the interesting features of the geometry.
In section~\ref{sec:URQMDfreezeout}, a realistic transport model (\UrQMD) is used to generate a freezeout distribution
  featuring non-Gaussian geometry and strong collective flow.
We discuss the non-trivial anisotropies of the distribution and the relationship between regions of homogeneity~\cite{Makhlin:1987gm,Akkelin:1995gh}
  and the ``whole'' source of emission points.
In section~\ref{sec:Elliot}, we simulate an experimental analysis, using the \UrQMD-generated distributions.
We build two-pion correlation functions and use the formulae presented in section~\ref{sec:formalism} to estimate
  the source anisotropies.
The calculations are compared to reported experimental results from Au+Au collisions at $\sqrt{s_{NN}}=3.84$~GeV, and predictions
  are given for heavy ion collisions at $\sqrt{s_{NN}}=30$~GeV, relevant for FAIR and the RHIC energy scan.
In section~\ref{sec:Summary}, we summarize.

\section{HBT radii and their connection to the underlying source}
\label{sec:formalism}

Femtoscopic two-particle correlation functions as a function of the relative momentum $q=p_1-p_2$
  are often fitted in terms of Gaussian HBT radii $R^2_{i,j}$
\begin{equation}
C\left(q\right) = 1+\lambda\exp\left(-\sum_{i,j=o,s,l}q_iq_jR^2_{i,j}\right) .
\label{eq:GaussianFit}
\end{equation}
Indices $i$ and $j$ indicate the components of the $q$ vector in the so-called Bertsch-Pratt ``out-side-long'' coordinate
  system~\cite{Pratt:1986cc,Bertsch:1989vn,Csorgo:1989kq}.
Here, ``out'' points along the direction of the
  pair transverse momentum,  and ``long'' points along the direction of motion of one of the incoming nuclei, say, in
  the direction of Au ions in the yellow ring of RHIC.
We shall call this the ``yellow nucleus'' and its colliding partner the ``blue nucleus.''
(We return to this arbitrary designation soon.)
The ``side'' direction is given by the cross-product of ``out'' with ``long.''
It is worthwhile to point out that, while 
  one may simultaneously flip the signs of all components of $q$ by swapping the designation of particles~1 and~2,
  the correlation function depends only on even-order products of $q$'s components; these products have meaningful sign.

\begin{figure}[t!]
{\centerline{\includegraphics[width=0.45\textwidth]{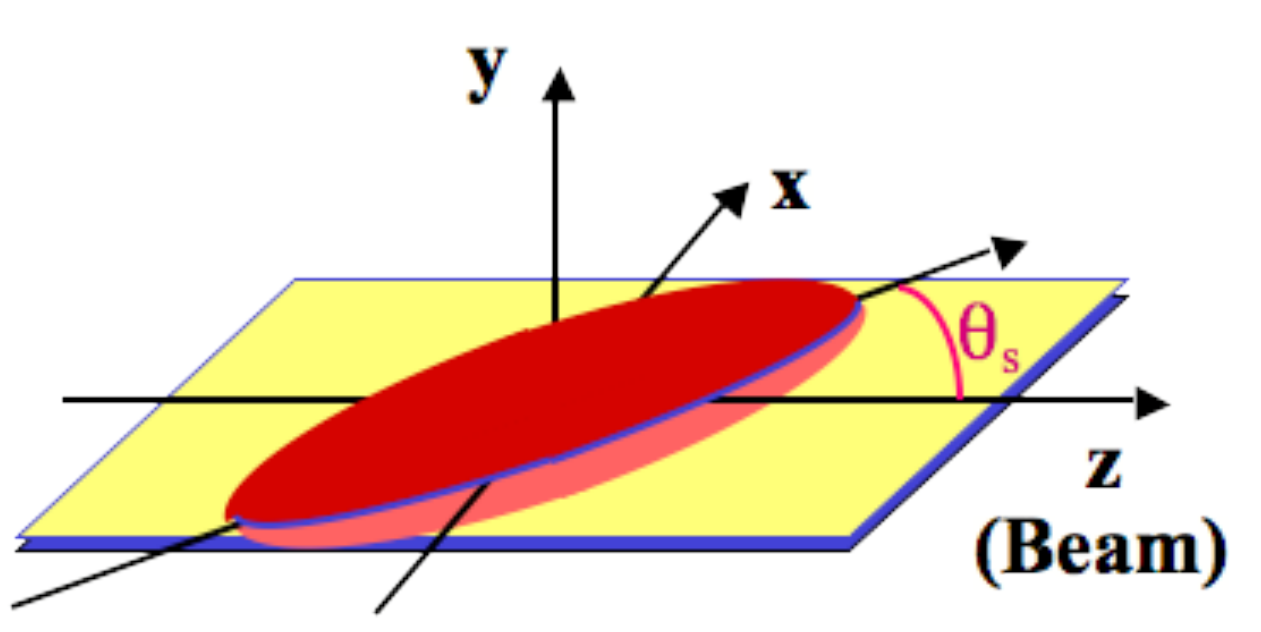}}}
\caption{(color online) The simplified parameterization of the freezeout distribution in a heavy ion
collision.  In addition to the timescale and three spatial length scales, the ellipsoid may be tilted
in the direction of the impact parameter $x$, relative to the beam axis $z$.
\label{fig:TiltedEllipseCartoon}
}
\end{figure}

We begin by considering a simple source of midrapidity pions which is a Gaussian ellipsoid in space and time, with the major axis
  of the ellipse tilted with respect to the beam direction, as sketched in Figure~\ref{fig:TiltedEllipseCartoon}.
The distribution is characterized by five parameters: a temporal scale, three spatial scales and a tilt angle
\begin{align}
f\left(x,y,z\right) \sim \exp\left( -\frac{\left(x\cos\theta_s-z\sin\theta_s\right)^2}{2\sigma_{x^\prime}^2} \right.
                -\frac{y^2}{2\sigma_y^2}   \notag \\
                \left. -\frac{\left(x\sin\theta_s+z\cos\theta_s\right)^2}{2\sigma_{z^\prime}^2}
                -\frac{t^2}{2\sigma_t^2}\right) ,
\label{eq:GaussianSource}
\end{align}
where the primes on $\sigma_{x^\prime}$ and $\sigma_{z^\prime}$ signify that these correspond to principle axes of the ellipse.


Its transverse eccentricity about its (tilted) major axis is defined as
\begin{equation}
\epsilon^\prime \equiv \frac{\sigma^2_y-\sigma^2_{x^\prime}}{\sigma^2_y+\sigma^2_{x^\prime}}
\label{eq:epsPrime}
\end{equation}

If $\theta_s=0$ and $\sigma_x = \sigma_y \equiv \sigma_\perp$ (or equivalently in an azimuthally-integrated analysis),
  only three parameters characterize the source, and the HBT radii are given by
\begin{align}
R^2_s = \sigma^2_\perp  \nonumber \\
R^2_o = \sigma^2_\perp + \beta_\perp^2\sigma^2_t \\
R^2_l = \sigma^2_z + \beta_l^2\sigma^2_t\, , \nonumber
\end{align}
where $\beta_\perp$ and $\beta_l$ are the transverse and longitudinal velocities of the pion pair.
``Cross-term'' radii $R^2_{i\neq j}$ vanish by symmetry~\cite{Lisa:2000ip,Heinz:2002au}.

For the more general case, there are six HBT radii, and they depend on the azimuthal angle $\phi$.
This angle is meaningfully defined over the range [0,2$\pi$] about the beam direction relative the impact parameter,
  which is defined as the direction perpendicular to the beam, pointing {\it from} the yellow nucleus {\it to} the blue one.
Swapping the designations of the ``yellow'' and ``blue'' nuclei reverses the direction of the impact parameter ($x$ direction);
  however, it also reverses the ``long'' ($z$) direction, defined earlier.
Hence, the sense of the tilt $\theta_S$ is well-defined; a source with positive tilt features a positive
  spatial $x-z$ correlation, as shown in figure~\ref{fig:TiltedEllipseCartoon}.
Experimental measurement of the sense of the tilt~\cite{Lisa:2000xj} requires measuring the direction of the impact parameter,
  for example through directed flow of net baryons at forward rapidity.

The HBT radii, measured as a function of angle $\phi$ relative to the beam axis, are driven by source widths $\sigma_x,\sigma_y,\sigma_z$,
  rather than $\sigma_{x^\prime},\sigma_y,\sigma_{z^\prime}$,
The relationships between these widths are given by
\begin{align}
\sigma^2_{x^\prime} = \sigma^2_x\cos^2\theta_s + \sigma^2_z\sin^2\theta_s + \sigma^2_{xz}\sin2\theta_s \notag \\
\sigma^2_{z^\prime} = \sigma^2_x\sin^2\theta_s + \sigma^2_z\cos^2\theta_s - \sigma^2_{xz}\sin2\theta_s 
\end{align}
where $\sigma^2_{xz}$ is the covariance between $x$ and $z$ in the source function; see~\cite{Lisa:2000ip} for
  details.

The HBT radii are related to these widths as~\cite{Lisa:2000ip}
  \begin{eqnarray}
    && R_s^2\left(\phi\right) = \textstyle{1\over 2} \left(\sigma^2_x + \sigma^2_y\right)
             + \textstyle{1\over 2} \left(\sigma^2_y - \sigma^2_x\right)
               \cos 2\phi\, ,
  \nonumber \\
    && R_o^2\left(\phi\right) = \textstyle{1\over 2} \left(\sigma^2_x + \sigma^2_y\right)
             - \textstyle{1\over 2} \left(\sigma^2_y - \sigma^2_x\right) \cos 2\phi\,
             + \beta_\perp^2 \sigma^2_t\, ,
  \nonumber \\
    && R_{os}^2\left(\phi\right) = \textstyle{1\over 2} \left(\sigma^2_y-\sigma^2_x\right) \sin 2\phi
  \nonumber \\
    && R_{l}^2\left(\phi\right) = \sigma^2_z + \beta_l^2 \sigma^2_t \, ,
  \nonumber \\
    && R_{ol}^2\left(\phi\right) = \sigma^2_{xz} \cos\phi\, ,
  \nonumber \\
    && R_{sl}^2\left(\phi\right) = - \sigma^2_{xz} \sin\phi\,  .
%
  \label{eq:osc1}
  \end{eqnarray}

In analogy with Equation~\ref{eq:epsPrime}, we identify the eccentricity of the source around the {\it beam} axis as
\begin{equation}
\epsilon \equiv \frac{\sigma^2_y-\sigma^2_{x}}{\sigma^2_y+\sigma^2_{x}}
\label{eq:eps}
\end{equation}

Equations~\ref{eq:osc1} express the explicit $\phi$-dependence of the HBT radii for a non-flowing source; all other variables in
  the equations are constants, for a non-flowing source.
For a source with flow, the constants, e.g. $\sigma_y$, may themselves depend on $\phi$.
In this case, our equations will be violated; below, we quantify this violation and its effect on the extraction of $\theta_s$ and $\epsilon$.

Experimentally, one measures the squared HBT radii and calculates the Fourier coefficients quantifying their azimuthal dependence,
  per~\cite{Heinz:2002au,Retiere:2003kf}
\begin{eqnarray}
   R_s^2(\phi) &=& R_{s,0}^2 + {\textstyle2\sum_{n=2,4,6,\dots}}
   R_{s,n}^2\cos(n\phi),
 \nonumber\\
   R_o^2(\phi) &=& R_{o,0}^2 + {\textstyle2\sum_{n=2,4,6,\dots}}
   R_{o,n}^2\cos(n\phi),
 \nonumber\\
   R_{os}^2(\phi) &=&
   {\textstyle2\sum_{n=2,4,6,\dots}} R_{os,n}^2\sin(n\phi), 
  \\
   R_l^2(\phi) &=& R_{l,0}^2 + {\textstyle2\sum_{n=2,4,6,\dots}}
   R_{l,n}^2\cos(n\phi),
  \nonumber \\
  R_{ol}^2(\phi) &=& {\textstyle2\sum_{n=1,3,5,\dots}}R_{ol,n}^2\cos\phi
  \nonumber \\
  R_{sl}^2(\phi) &=& {\textstyle2\sum_{n=1,3,5,\dots}}R_{sl,n}^2\sin\phi .
  \nonumber 
\label{eq:FourierDecomposition}
\end{eqnarray}
Equivalently,
\begin{equation}
\label{eq:HBTradii-FCs}
R^2_{\mu,n}(p_T) =
\begin{cases}
\langle R^2_\mu(p_T,\phi_p) \cos(n\phi_p) \rangle & (\mu = o, s, l, ol) \\
\langle R^2_\mu(p_T,\phi_p) \sin(n\phi_p) \rangle & (\mu = os, sl)
\end{cases}.
\end{equation}

In our no-flow Gaussian model, then, the source geometry and orientation are extracted from the Fourier coefficents.
Identifying the tilt angle requires~\cite{Lisa:2000ip,Heinz:2002au} measuring HBT relative to the first-order
  reaction plane~\cite{Voloshin:2008dg}.
Published results from the STAR~\cite{Adams:2003ra} and CERES~\cite{Adamova:2008hs} collaborations use only the $2^{\rm nd}$-order
  plane and so report the source eccentricity only about the beam axis.
(Efforts to perform the analysis relative to the first-order plane are underway at RHIC~\cite{Aggarwal:2010cw}.)
In this case~\cite{Retiere:2003kf}
\begin{equation}
\epsilon = 2\cdot\frac{\tilde{R}^2_{s,2}}{R^2_{s,0}} .
\label{eq:ExtractingEpsilon}
\end{equation}
For the moment, we ignore the tildes above the $n\neq 1$ Fourier coefficients here and below.
They represent a trivial finite-binning correction discussed later in section~\ref{sec:Elliot}.

If the first-order plane is identified, first-order azimuthal oscillations in $R^2_{sl}$ and $R^2_{ol}$ are measureable.
In this case one obtains the tilt angle according to~\cite{Lisa:2000ip}
\begin{equation}
\theta_s = \textstyle{1\over 2} \tan^{-1}\left(\frac{-4\tilde{R}^2_{sl,1}}{R^2_{l,0}-R^2_{s,0}+2\tilde{R}^2_{s,2}}\right) .
\label{eq:ExtractingTheta}
\end{equation}
The transverse eccentricity in the ``natural'' frame tilted relative to the beam axis is
\begin{equation}
\epsilon^\prime = \frac{2\tilde{R}^2_{s,2}\left(1+\cos^2\theta_s\right)+\left(R^2_{s,0}-R^2_{l,0}\right)\sin^2\theta_s-2\tilde{R}^2_{sl,1}\sin2\theta_s}
                       {R^2_{s,0}\left(1+\cos^2\theta_s\right)+\left(2\tilde{R}^2_{s,2}+R^2_{l,0}\right)\sin^2\theta_s+2\tilde{R}^2_{sl,1}\sin2\theta_s}
\label{eq:ExtractingEpsilonPrime}
\end{equation}

\section{Freezeout distributions from a more realistic model}
\label{sec:URQMDfreezeout}

Our simple model of the emission function given in equation~\ref{eq:GaussianSource} is unrealistic in at least two ways.
Firstly, while realistic emission functions are often roughly Gaussian, they are never perfectly so; in this case, the two-pion
  correlation function is likewise non-Gaussian.
Extracting Gaussian HBT errors through fits with equation~\ref{eq:GaussianFit}, then, could in principle generate considerable
  mis-representation of the emission function.

The second over-simplification of the source discussed above is its lack of collective flow, which generates correlations between
  a particle's emission position and its momentum~\cite{Makhlin:1987gm}.
For example, an explosively flowing source will boost particles emitted from its right side, towards the right.
The freezeout distribution of particles with a given momentum vector is known as the region of homogeneity for that momentum vector.
In {\it principle}, the homogeity regions for different azimuthal angles might be completely disjoint and unrelated, obviously invalidating
  equations~\ref{eq:osc1}, \ref{eq:ExtractingEpsilon}, \ref{eq:ExtractingTheta} and~\ref{eq:ExtractingEpsilonPrime}.
In practice, the homogeneity regions in blast-wave models~\cite{Retiere:2003kf} or hydodynamic simulations~\cite{Heinz:2002sq} are
  naturally related.
For boost-invariant sources, equation~\ref{eq:ExtractingEpsilon} is estimated to be good to $\sim 30\%$ in these models~\cite{Retiere:2003kf}.

While blast-wave and boost-invariant hydro models do feature non-Gaussian sources and collective flow, they are still simplistic.
Firstly, any boost-invariant model by definition has no tilt relative to the beam axis; thus they are unable to access physics
  associated with $\theta_s$.
Secondly, they typically use the Cooper-Frye formula to model particle freezeout from a calculated or parameterized hypersurface;
  while momentum-space observables (e.g. $v_2$) may be insensitive to this procedure, interferometry is known to be quite sensitive
  to the freezeout proceedure.

In this section, we use the Ultra-Relativistic Quantum Molecular Dynamics Model (\UrQMD~3.3)\footnote{This version can be downloaded from
  www.urqmd.org.} to generate a realistic freeze-out distribution with fully three-dimensional dynamics~\cite{Bass:1998ca,Bleicher:1999xi}.
\UrQMD~ is a
covariant transport approach to simulate the interactions between hadrons
and nuclei up to relativistic energies. It is based on the propagation of
nucleons and mesons accompanied by string degrees of freedom with
interaction probabilities according to measured and calculated cross
sections for the elementary reactions. Hard scatterings with large momentum
transfer are treated via PYTHIA.  For detailed comparisons of this version to
experimental data, the reader is referred to \cite{Petersen:2008kb}.
Previous HBT studies with \UrQMD~ have been reported
  in~\cite{Li:2006gp,Li:2006gb,Li:2007im,Li:2008qm}.


\begin{figure}[t!]
{\centerline{\includegraphics[width=0.32\textwidth]{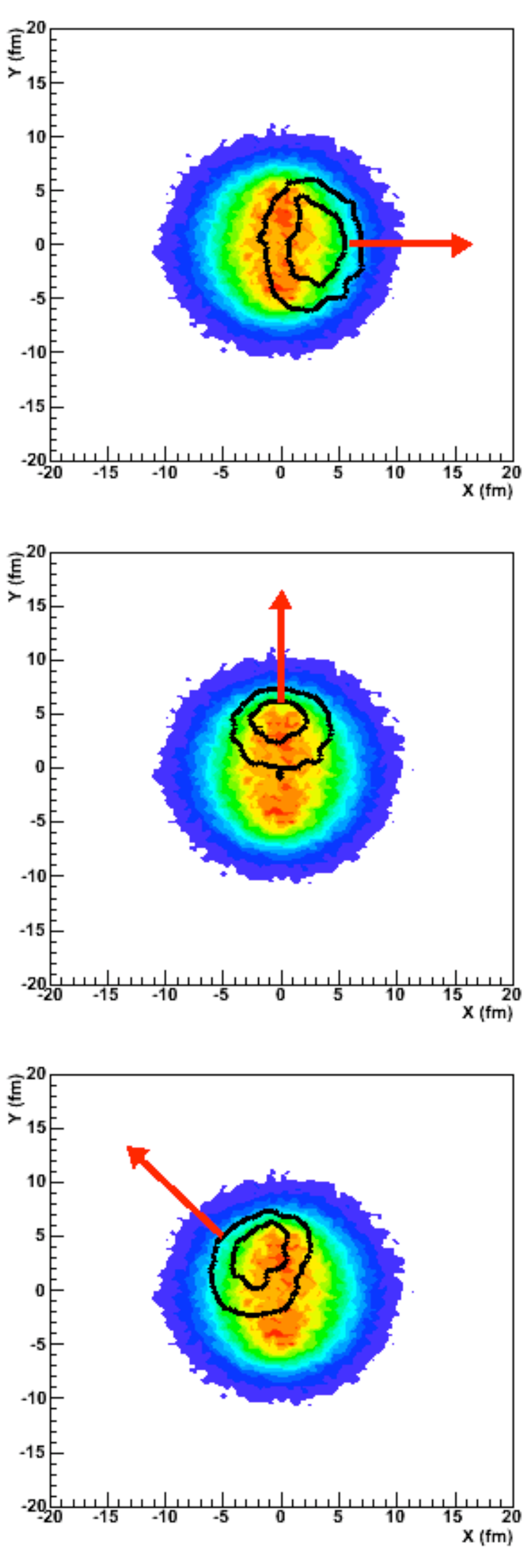}}}
\caption{(color online) Pion emission points from \UrQMD~ simulations of Au+Au collisions with collision energy $\sqrt{s_{NN}}=3.84$~GeV
  (corresponding to a 6~AGeV beam incident on a fixed target) and impact parameter $b=4-8$~fm.
  Colored contours (identical for the three panels) show the emission point density of all pions with $p_T<0.4$~GeV/c.  Black contours in the upper, middle and lower panels
  indicate the density of emission points for pions with $\phi=\left[-\tfrac{\pi}{8},\tfrac{\pi}{8}\right]$, $\left[\tfrac{3\pi}{8},\tfrac{5\pi}{8}\right]$ and
  $\left[\tfrac{5\pi}{8},\tfrac{7\pi}{8}\right]$, respectively.
\label{fig:HomogeneityRegions}
}
\end{figure}

\begin{figure}[t!]
{\centerline{\includegraphics[width=0.45\textwidth]{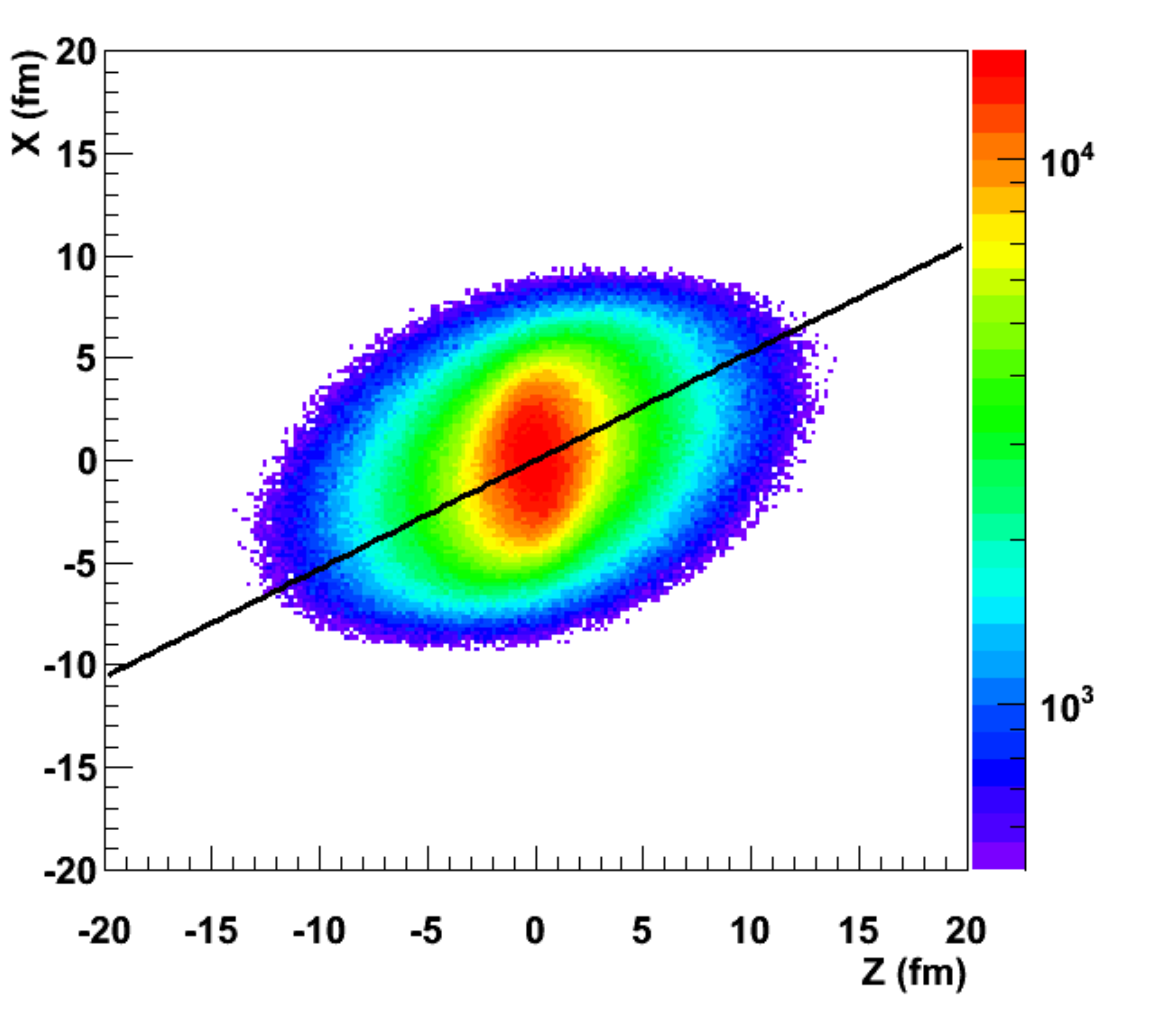}}}
\caption{Freezeout distribution of pions from $\sqrt{s_{NN}}=3.84$~GeV (6~AGeV beam energy on fixed target)
Au+Au collisions with impact parameter $b=4-8$~fm in the reaction
  ($x-z$) plane, as calculated from \UrQMD.
}
\label{fig:FreezeoutInRP}
\end{figure}

In \UrQMD, the freeze-out space-time point is naturally defined as the last hard interaction of a particle.
The freezeout distribution of pions from $\sqrt{s_{NN}}=3.84$~GeV Au+Au collisions with impact parameter $b=4-8$~fm in the reaction plane
  ($x-z$ plane) is shown in figure~\ref{fig:FreezeoutInRP}.  The source has an obvious tilt structure relative to the beam axis.
We use the parameterization of equation~\ref{eq:GaussianSource} to fit the three-dimensional freezeout distribution for {\it all}
  pions with $p_T<400$~MeV/c and $|y|<0.6$-- not only those at a given angle $\phi$.
From this direct analysis of the freezeout coordinates-- obviously not possible in experiment-- we find the parameters listed
  in the third column of table~\ref{tab:Parameters}.
It is clear from figure~\ref{fig:FreezeoutInRP} that $x-z$ correlation in the freezeout distribution has structure that cannot
  be captured in a single tilt number; indeed, the tilt is scale-dependent, growing as one focuses on the peak of the distribution.
This ``twist'' feature, which has been observed in simulations before~\cite{Lisa:2000ip}, might be physically interesting and
  experimentally accessible; we leave exploration of this effect for a future work.
For the purpose of this work, we identify a range of tilts arising from fitting equation~\ref{eq:GaussianSource} to the distribution
  and varying the fit range in coordinate space from $10~{\rm fm}<\Delta x,\Delta y,\Delta z<40~{\rm fm}$.
This leads to the range shown in the left column of table~\ref{tab:Parameters}.

Dynamics naturally leads to a strong correlation between a particle's final momentum and the freezeout position; homogeneity regions
  are naturally reflected in the final state.
Figure~\ref{fig:HomogeneityRegions} shows homogeneity regions from UrQMD~ in the $x-y$ plane for particles emitted at three azimtuthal angles.
The homogeneity region for particles emitted at a given angle $\phi_1$ clearly differs from that for particles emitted at $\phi_2\neq\phi_1$.
Thus, in addition to the {\it explicit} $\phi$-dependence of the HBT radii seen in equations~\ref{eq:osc1}, there is an additional
  {\it implicit} dependence~\cite{Voloshin:1996ch,Wiedemann:1997cr,Heiselberg:1998ik,Heiselberg:1998es}.
HBT radii measured at a given momentum vector $\left(\phi,p_T,y\right)$ probe only the geometry of the homogeneity region for that momentum vector.
{\it A priori}, it is far from clear that equations~\ref{eq:ExtractingEpsilon}, \ref{eq:ExtractingTheta} and~\ref{eq:ExtractingEpsilonPrime}, which attempt to relate HBT radius
  oscillations to the geometry of the ``whole'' source, will prove good approximations.

In the next section, we test these relations-- true for the simplest toy model of Equation~\ref{eq:GaussianSource}-- with \UrQMD-generated freezeout distributions.

\section{HBT radii and anisotropy parameters from UrQMD-generated correlation functions}
\label{sec:Elliot}

We start by generating two-pion correlation functions, analogous to those measured experimentally, from \UrQMD~ events.
We will then proceed to fit these correlation functions with the Gaussian anzatz of equation~\ref{eq:GaussianFit}, as is done in experimental analysis.
Finally, we extract Fourier coefficients characterizing the oscillations of the HBT radii with angle; from these
  we extract the source anisotropy parameters that would be obtained in an experiment.

\begin{figure}[t!]
{\centerline{\includegraphics[width=0.44\textwidth]{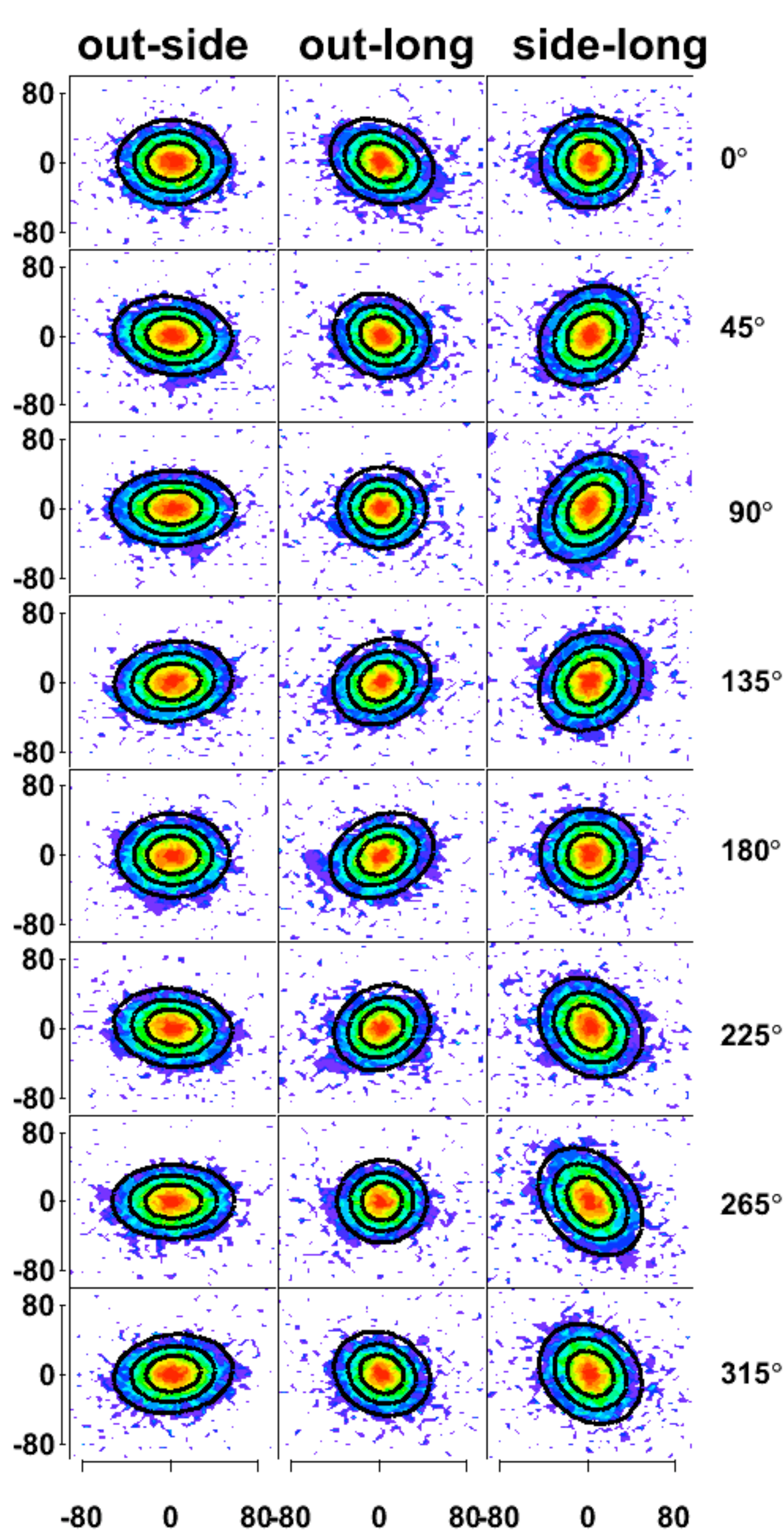}}}
\caption{(color online) Pion correlation functions from Au+Au collisions with collision energy $\sqrt{s_{NN}}=3.84$~GeV and 
  (corresponding to a 6~AGeV beam incident on a fixed target) and impact parameter $b=4-8$~fm, as calculated with the \UrQMD~ model.
  Projections in the $q_o-q_s$ (left column), $q_o-q_l$ (middle) and $q_s-q_l$ spaces are made, with the unplotted $\vec{q}$-component
  smaller than 4~MeV/c.  Correlation slices are shown for pion pairs in $45^\circ$-wide $\phi$ bins centered at angles indicated to the right
  of each row.
  Shaded (color online) contours represent the calculated correlation function.  Black contours represent 2-dimensional slices of the
  three-dimensional Gaussian fit to the correlation functions for each selection in $\phi$.
\label{fig:2dCFs}
}
\end{figure}

To simulate experimental conditions, two-pion correlation functions were constructed
  from the \UrQMD~ events using the so-called weighting method~\footnote{This is discussed further in section 2.8 of~\cite{Lisa:2005dd}, where
  it is called Method II.}.
In this method, pairs of identical pions are selected according to a Monte Carlo algorithm;
  the correlation function in a given $\left(q_o,q_s,q_l\right)$ bin is equal to the
  pair-wise average of the squared two-pion squared wavefunction.
Considering only quantum symmetrization effects, the correlation function is computed as
\begin{equation}
\label{eq:WeightingMethod}
C\left(q_o,q_s,q_l;\phi\right) = \langle 1+\cos\left(-q^\mu\Delta r_\mu\right) \rangle_\phi ,
\end{equation}
where $q=p_1-p_2$ is the relative momentum and $\Delta r$ is the space-time separation of the particles at freezeout.

As explicitly denoted in equation~\ref{eq:WeightingMethod}, 
  the correlation functions were generated for 8 $45^\circ$-wide bins in pair angle $\phi\equiv\angle(\vec{K}_T,\vec{b})$, where
  $\vec{K}_T\equiv\left(\vec{p}_{T,1}+\vec{p}_{T,2}\right)/2$ is the average transverse momentum vector of the pair.
Hence, for $-\tfrac{\pi}{8}<\phi<\tfrac{\pi}{8}$ ($\tfrac{5\pi}{8}<\phi<\tfrac{7\pi}{8}$), pions from sub-region indicated by the black contours in the top
  (bottom) panel of figure~\ref{fig:HomogeneityRegions} are used to construct the correlation function.

\begin{figure}[t!]
{\centerline{\includegraphics[width=0.5\textwidth]{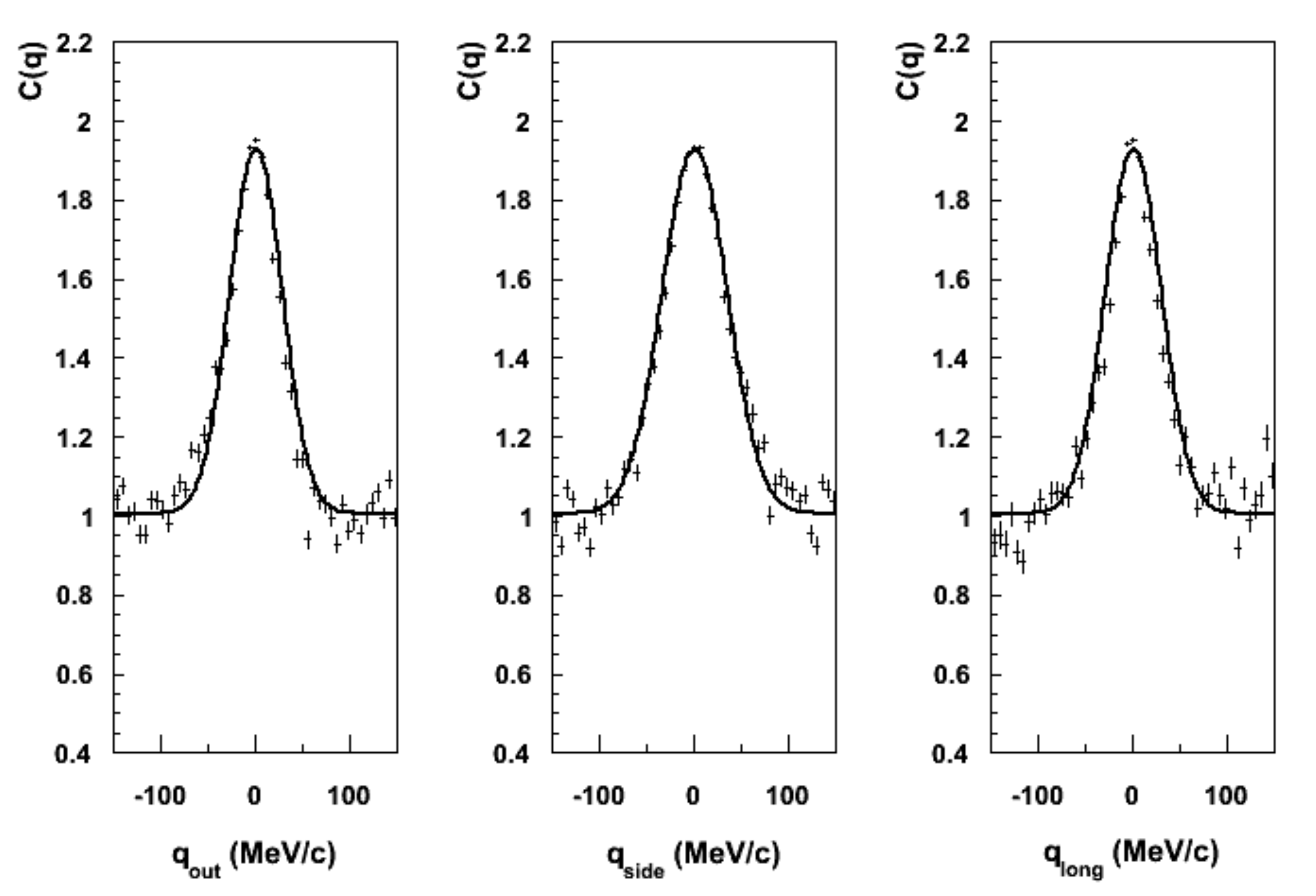}}}
\caption{One dimensional slices in the three components of relative momentum, for pions emitted at $|\phi|<22.5^\circ$.  Curves
represent one-dimensional slices of the three-dimensional Gaussian fit to the correlation function.
\label{fig:1dCF}
}
\end{figure}

In figure~\ref{fig:2dCFs} are plotted, for each bin in $\phi$,
   two-dimensional slices of the three-dimensional correlation function in the $q_o-q_s$, $q_s-q_l$ and $q_l-q_o$ planes; in each case the unplotted
  relative momentum component $q_i<4$~MeV/c.
For a representative $\phi$ bin, one-dimensional slices of the correlation function in the out, side, and long directions are
  shown in Figure~\ref{fig:1dCF}.
Most femtoscopic correlation analyses focus on the one-dimensional projections, since the correlation in three-dimensional space
  factorizes; that is, there is no covariance between components $q_i$ and $q_{j\neq i}$ in the correlation function.
This is not the case when the analysis is performed differentially in 
  $\phi$~\cite{Voloshin:1996ch,Wiedemann:1997cr,Heiselberg:1998ik,Heiselberg:1998es,Lisa:2000ip,Lisa:2000xj,Heinz:2002au},
  as is clear from the tilted
  structures in $\vec{q}$-space seen in figure~\ref{fig:2dCFs}.
These tilts in the individual correlation functions in $\vec{q}$-space for a given angular selction in $\vec{K}$ are not to
  be confused with the overall spatial tilt of the source sketched in figure~\ref{fig:TiltedEllipseCartoon}.

\begin{figure}[t!]
{\centerline{\includegraphics[width=0.5\textwidth]{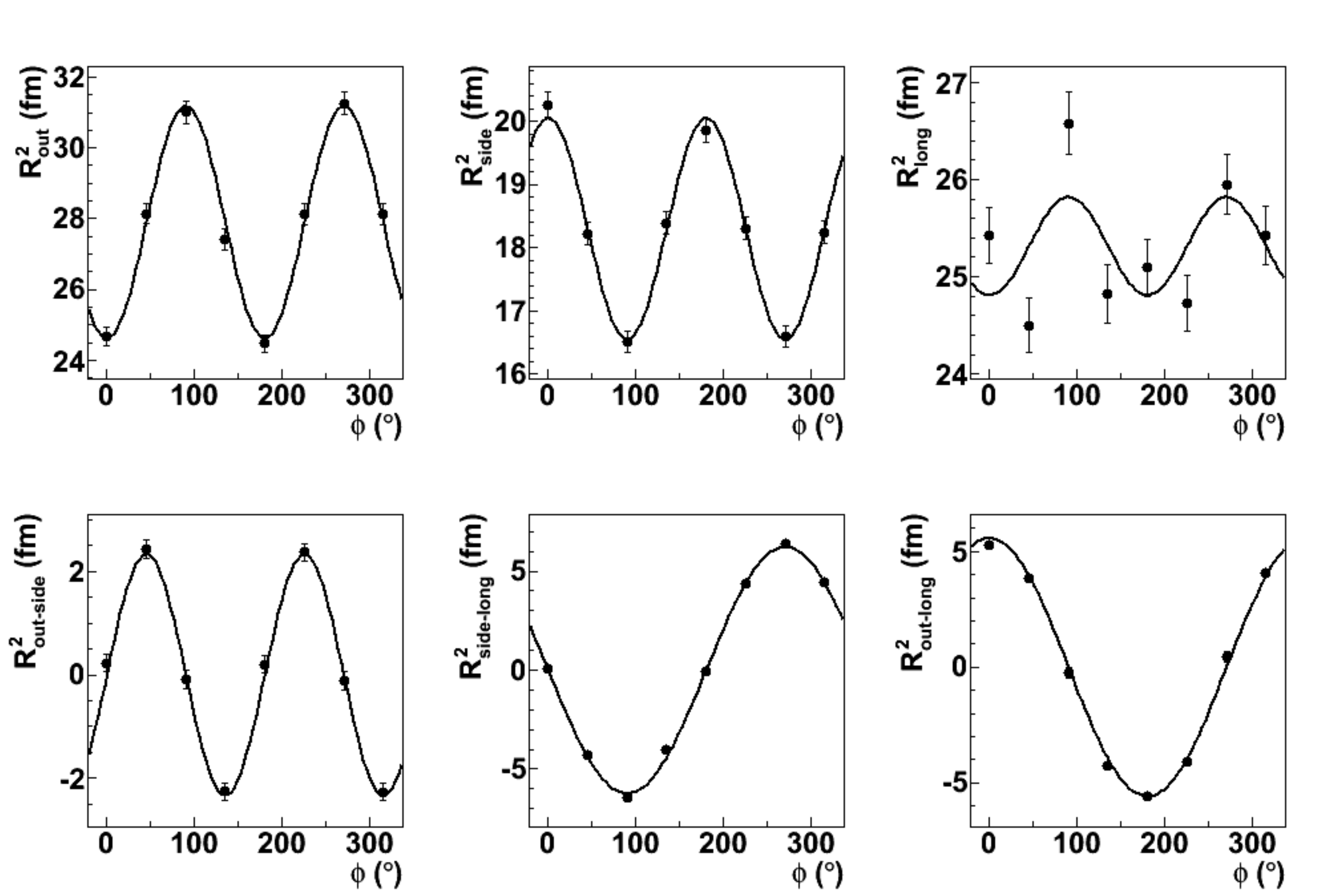}}}
\caption{The six HBT radii extracted from Gaussian fits (equation~\ref{eq:GaussianFit}) to the correlation functions for eight selections
  on $\phi$.
  Curves represent a Fourier decomposition (equation~\ref{eq:FourierDecomposition}), including terms up to second order, where the Fourier
  components are determed according to equation~\ref{eq:HBTradii-FCs}.
  See text for details.
\label{fig:HBTradii}
}
\end{figure}

\begin{figure}[h]
{\centerline{\includegraphics[width=0.5\textwidth]{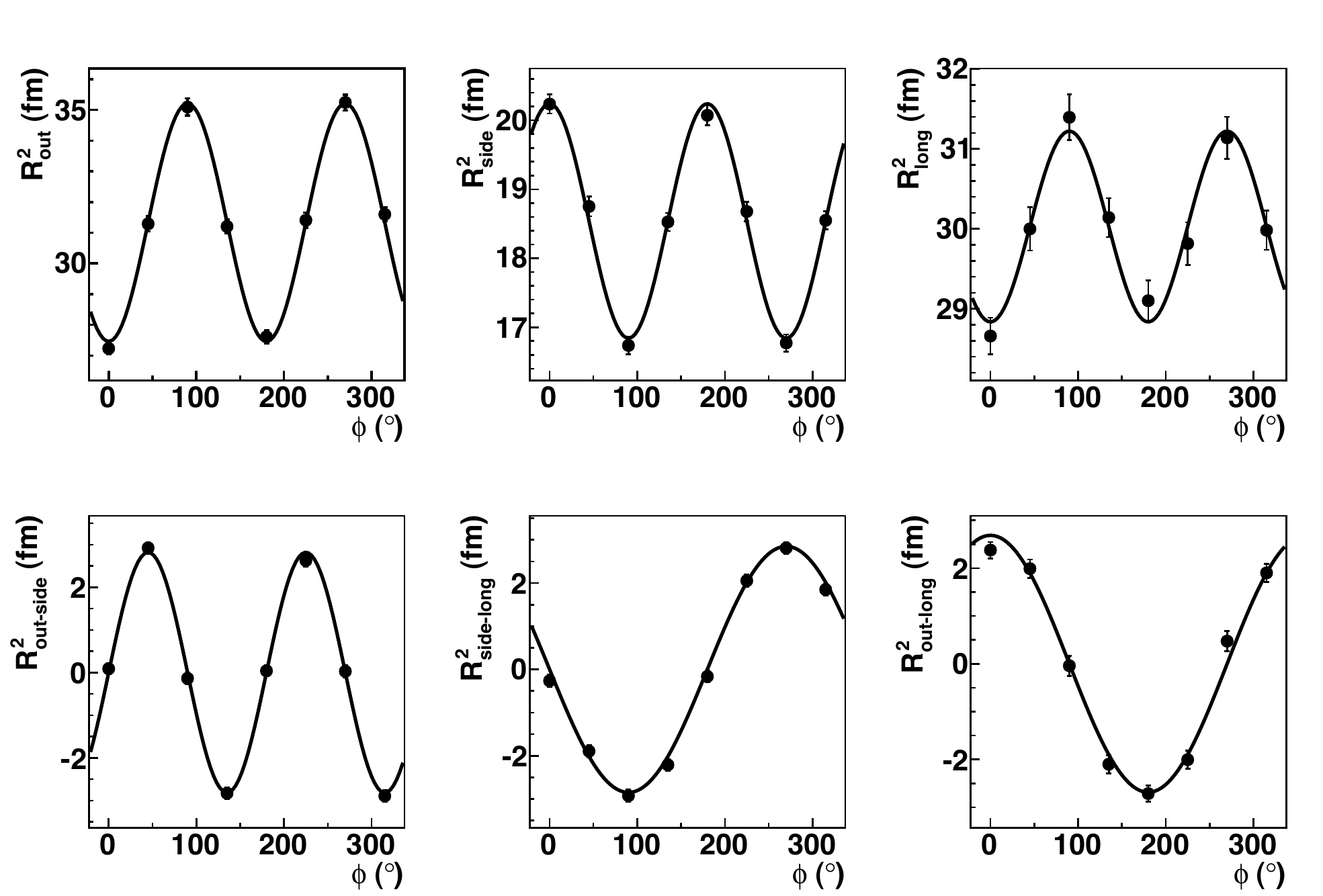}}}
\caption{Same as figure~\ref{fig:HBTradii}, but for collisions at $\sqrt{s_{NN}}=7.7$~GeV.
\label{fig:HBTradii30GeV}
}
\end{figure}

As in an experimental analysis, the correlation functions are fitted with the Gaussian functional form of equation~\ref{eq:GaussianFit}.
Two- and one-dimensional slices of these fits are superimposed on the correlation functions in figures~\ref{fig:2dCFs} and~\ref{fig:1dCF}.
The six resulting HBT radii are plotted as a function of $\phi$ on figure~\ref{fig:HBTradii}.
As in a standard azimuthally-integraged analysis, the ``diagonal'' radii $R^2_{j},\, j=o,s,l$ are driven by the width of the correlation
  function in the direction $j$.
The ``cross-term'' radii $R^2_{i,j\neq i}$ quantify the correlations between $\vec{q}$ components-- the tilts of the correlation function;
  e.g. the $\phi$-dependence of the tilt of the correlation function in the $q_s-q_l$ space seen in the right column of figure~\ref{fig:2dCFs}
  leads directly to the first-order oscillation of $R^2_{sl}$ seen in figure~\ref{fig:HBTradii}.
Figure~\ref{fig:HBTradii30GeV} shows the radii for $\sqrt{s_{NN}}=7.7$~GeV collisions.

The curves on figures~\ref{fig:HBTradii} and~\ref{fig:HBTradii30GeV} represent equations~\ref{eq:FourierDecomposition} using
  Fourier coefficients $R^2_{\mu,n}$ calculated according to~\ref{eq:HBTradii-FCs}.

Before we apply equations~\ref{eq:ExtractingEpsilon}-\ref{eq:ExtractingEpsilonPrime}, we note that
  the argumentation of section~\ref{sec:formalism} 
  implicitly assumed that the $R^2_{\mu}\left(\phi\right)$ were measured as a continuous function.  
In reality, in our analysis as in experiment, the correlation function is measured for $\phi$ within bins of width $\Delta\phi$.
The amplitude of the $n^{\rm th}$-order oscillation of a binned function is reduced from that of the underlying function.
To correct for this finite-binning artifact, we calculate the underlying (``true'') Fourier coefficients $\tilde{R}^2_{\mu,n}$ from
  the ones extracted from the binned radii by
\begin{align}
\tilde{R}^2_{\mu,n} = \frac{n\Delta\phi/2}{\sin\left(n\Delta\phi/2\right)}R^2_{\mu,n}\, .
\label{eq:BinningCorrection}
\end{align}
For our $45^\circ$-wide bins, $\tilde{R}^2_{\mu,1}=1.026R^2_{\mu,1}$ and $\tilde{R}^2_{\mu,2}=1.111R^2_{\mu,2}$.
These binning-corrected Fourier coefficients are used in equations~\ref{eq:ExtractingEpsilon}-\ref{eq:ExtractingEpsilonPrime}.
 \footnote{In principle, one could correct the correlation functions themselves for the finite $\phi$-binning, and then extract
           HBT radii from them, as described in~\cite{Heinz:2002au}.  However, especially if the reaction-plane estimation
           resolution~\cite{Voloshin:2008dg} is good (in our model analysis, it is perfect), it makes no significant difference
           whether the correlation functions or the radii themselves are corrected for binning effects~\cite{Wells:2002phd}.}

\begin{figure}[t!]
{\centerline{\includegraphics[width=0.4\textwidth]{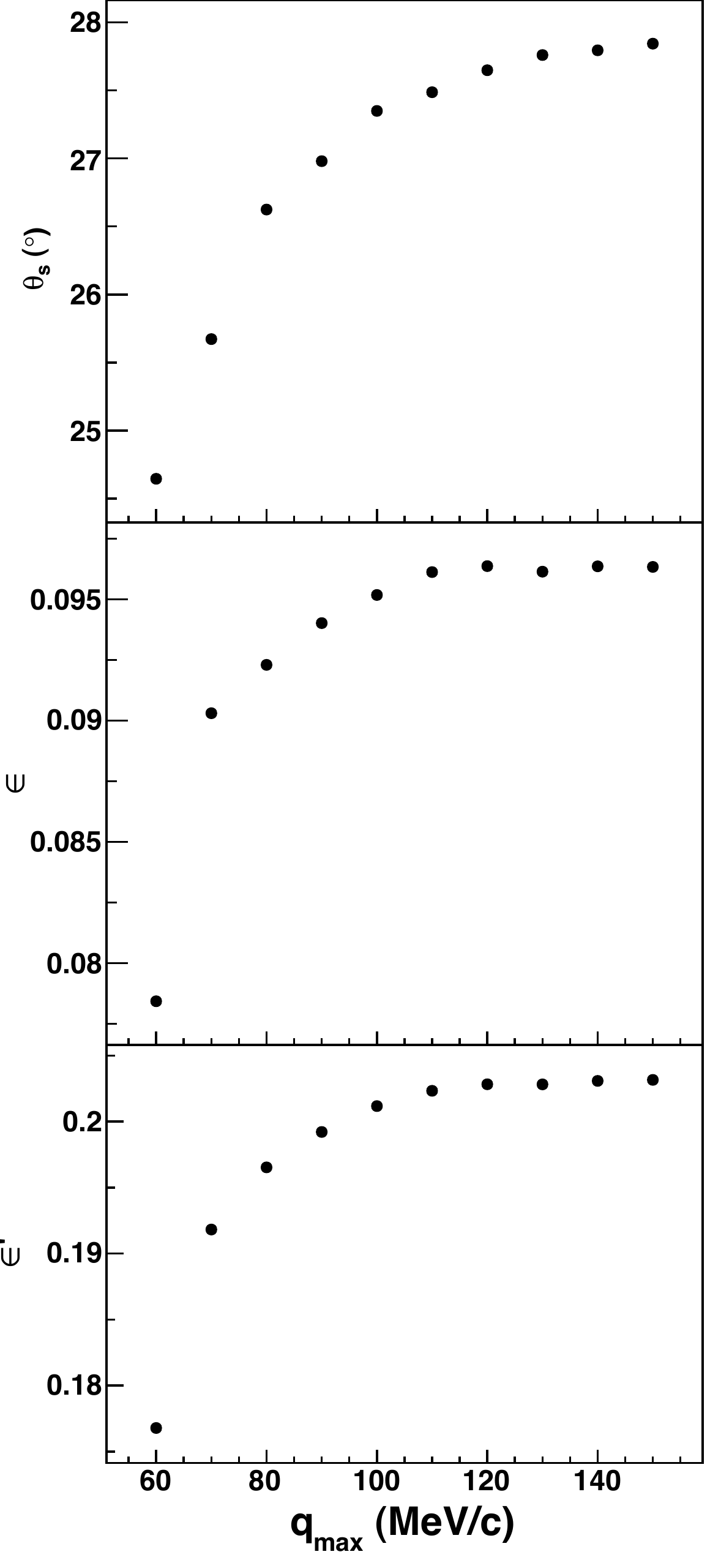}}}
\caption{Source anisotropy parameters extracted from two-pion correlation functions for \UrQMD-generated Au+Au collisions
  at $\sqrt{s_{NN}}=3.84$~GeV at $b=4-8$~fm, are plotted as a function of the range in relative momentum over which the
  correlation functions are fitted.
  Respectively, top, middle and bottom panels show the source tilt angle $\theta_S$, eccentricity about the beam axis $\epsilon$ and
  eccentricity about the tilted axis $\epsilon^\prime$.
  See text for details.
\label{fig:FitRangeStudy}
}
\end{figure}

As in most experimental analyses, the correlation functions from \UrQMD~ simulations are not purely Gaussian,
  since the source itself is not Gaussian in coordinate space, due to space-momentum correlations (flow), resonance
  contributions, etc.
Of special interest for the present study is the additional fact that the source is not characterized by a tilt angle
  independent of spatial scale-- the ``twist'' discussed in section~\ref{sec:URQMDfreezeout}.
Following a standard experimental approach~\citep[e.g.][]{Adams:2004yc}, we perform a fit-range study,
  in which we vary the range in $q_o$, $q_s$, $q_l$,
  over which we perform the fit with equation~\ref{eq:GaussianFit}.
In particular, we fit the correlation functions in the range $-q_{max}<q_o,q_s,q_l<q_{max}$ for $q_{max}=60-150$~MeV/c.
The resulting anisotropy parameters for $\sqrt{s_{NN}}=3.84$~GeV collisions, calculated according to
  equations~\ref{eq:ExtractingEpsilon}-\ref{eq:ExtractingEpsilonPrime},
  are shown on figure~\ref{fig:FitRangeStudy}.
The dependence of $\theta_S$ on $q_{max}$ is readily understood.
Large values of $q$ probe smaller values of coordinate space; thus, as $q_{max}$ is increased, the fit
  is increasingly sensitive to the large tilt structure seen at small scales in figure~\ref{fig:FreezeoutInRP}.
Figure~\ref{fig:FitRangeStudy}, then, is itself a measure of the ``twist'' structure in coordinate space,
  though there may be more sophisticated ones.
For our purposes, however, we take the variation of the anisotropy parameters plotted in figure~\ref{fig:FitRangeStudy}
  to define a range of values one might expect from an experimental study.
Since a typical experimental analysis would fit at least out to 100~MeV/c (in order to include all of the peak signal), the
  value ranges listed in the fourth column of table~\ref{tab:Parameters} correspond to $100~{\rm MeV/c}<q_{max}<150~{\rm MeV/c}$.

The agreement with the parameters extracted via direct analysis of the \UrQMD~ freezeout coordinates is fair; and
  we discuss this further in the next section.

The fifth column in table~\ref{tab:Parameters} lists the anisotropy parameters based on HBT radii measured by the E895 collaboration
  in a fixed-target experiment at $E_{beam}=6$~AGeV~\cite{Lisa:2000xj}.
The \UrQMD~ calculation reproduces the tilt angle very well, the eccentricities somewhat less well, though experimental
  uncertainties are large.
\UrQMD~ calculations in the bottom row of the table represent a prediction for collisions at
  $\sqrt{s_{NN}}=7.7$~GeV, which will be measured at FAIR and RHIC.

\section{Discussion and Summary}
\label{sec:Summary}

\begin{table}[t!]
\begin{tabular}{|c|c|c|c|c|}
\hline
\multirow{2}{*}{$\sqrt{s_{NN}}$} & ~                          & Direct fit to                & UrQMD: HBT             & Expt.: HBT    \\
                                 & ~                          & coordinate space             & oscillations           & oscillations       \\
\hline
\multirow{1}{*}{3.84 GeV}        & ~~$\epsilon$~~             & $0.13-0.17$                  & $0.095-0.096$                & $0.30\pm0.15$      \\
          (6 AGeV)               & ~~$\theta_s$~~             & $34^\circ-41^\circ$          & $27.4^\circ-27.9^\circ$   & $26^\circ\pm7^\circ$ \\
                  ~              & ~~$\epsilon^\prime$~~      & $0.21-0.26$                  & $0.200-0.206$                 & $0.38\pm0.19$        \\
\hline
\multirow{1}{*}{7.7 GeV}         & ~~$\epsilon$~~             & $0.11-0.14$                  & $0.090-0.091$                  &  -                 \\
           (30 AGeV)             & ~~$\theta_s$~~             & $14^\circ-21^\circ$          & $11.7^\circ-11.9^\circ$                   &  -                 \\
                                 & ~~$\epsilon^\prime$~~      & $0.12-0.16$                  & $0.109$                   &  -                 \\
\hline
\end{tabular}
\caption{Source anisotropy parameters, for Au+Au collisions at two collision energies, with $b=4-8$~fm for pions with $p_T<0.6$~GeV/c, and $|y|<0.6$.
Impact parameter and momentum cuts were selected to match published data from the E895 collaboration~\cite{Lisa:2000xj}.
Third column: estimates from a Gaussian fit (equation~\ref{eq:GaussianSource}) to the freezeout distribution from \UrQMD~ events.
Fourth column: estimates using equations~\ref{eq:ExtractingEpsilon}-\ref{eq:ExtractingEpsilonPrime} on the azimuthal oscilations
   of HBT radii from \UrQMD~ events.
Fifth column: same as column four, but using experimental data from E895.
Experimental data at 7.7~GeV will be analyzed at FAIR and RHIC.}
\label{tab:Parameters}
\end{table}

The connection between anisotropic particle emission distributions and femtoscopic correlation functions has been discussed in detail.
Gaussian fits to three-dimensional two-pion correlation functions result in six HBT radii (in contrast to the usual three), all of which
  depend on the pair emission angle relative to the impact parameter.
The order and magnitude of the radius oscillations are quantified by Fourier coefficients, which may be directly
  related to the spatial tilt and eccentricity of a Gaussian source with no space-momentum correlations (arising, for example, from collective flow).
For a more realistic freezeout distribution, the connection between the oscillating radii and the source anisotropy is only approximate.
Using a sophisticated transport model to simulate the entire collision evolution and particle decoupling, we have studied
  the freezeout distribution in coordinate space.
The ``whole'' distribution indeed features a tilted structure and eccentricity, but shows other, less trivial anisotropic structures as well.

The experimentalist, of course, cannot study the distribution directly in coordinate space.
 We have simulated the experimental
  situation by constructing two-pion correlation functions for different pair emission angles, fitted each with a Gaussian functional
  form and extracted Fourier coefficients for each radius.
These coefficients were then used in the formalism derived in section~\ref{sec:formalism} to estimate the tilt and eccentricity.
The agreement between these estimates of the anisotropy and the more direct study in coordinate space, was fair.
Previous studies~\cite{Retiere:2003kf} of boost-invariant hydrodynamic and blast-wave models found that the eccentricity values
  estimated from two-particle radii were within 30\% of the ``true'' values; this has been used as a systematic error for the
  eccentricity in experimental studies~\cite{Adams:2003ra}.
Our results using \UrQMD~ are consistent with this 30\% value.
The present study is the first estimate of the corresponding uncertainty in the tilt, $\theta_S$;
  the agreement is somewhat worse, on the order of 35\%.
Given the complications of dynamically-induced homogeneity regions, a time-evolving emission distribution,
  non-Gaussianness and ``twist'' effects, one might easily have expected much worse agreement.

However approximate, quantifying the connection between the radius oscillations and the underlying geometry can be useful.
Ideally, the correct model of a heavy ion collision will reproduce all experimental observations; here, this means
  the HBT radii and their dependence on azimuthal angle.
However, when observations are reproduced and others not, connections to the underlying scenario are important.
For example, if a model reproduces $R_{long}$ and $R_{side}$ but over-estimates $R_{out}$~\citep[e.g.][]{Soff:2000eh,Soff:2002pc,Heinz:2002un},
  attention immediately turns to emission duration, which may be associated with the nature of the transition between
  confined and deconfined states, latent heat, etc~\cite{Bertsch:1989vn,Pratt:1986cc,Rischke:1996em}.

For the azimuthal dependence of HBT radii, the tilt angle and eccentricities probe different aspects of the dynamics.
At AGS energies ($\sim 3.5$~GeV), $\theta_S$ reflects the dynamics behind directed flow~\cite{Lisa:2000ip} in the earliest
  stage of the collision and shows strong dependence of the
  equation of state used in transport calculations~\cite{Lisa:2000xj}.
Meanwhile, the eccentricities represent the geometric and temporal~\cite{Lisa:2003ze} aspects of elliptic flow.
It is hoped that the connections we have discussed here, between experimental observations and the underlying
  source anisotropy, will help future comparative studies focus on the physics most relevant to each observable.

\section*{Acknowledgements}
This work supported by the U.S. National Science Foundation under Grant PHY-0970048 and by the Hessian
LOEWE initiative through HIC for FAIR and the Extreme Matter Institute EMMI. The authors would also like 
to acknowledge fruitful discussions with Dr. Hannah Petersen.  E.M. and M.A.L. gratefully acknowledge
the kind hospitality of the Frankfurt Institute for Advanced Studies.


\end{document}